\documentclass[aps,prb,twocolumn,superscriptaddress]{revtex4-1}

\usepackage{graphicx}

\begin{document}

\title{Evolution of London penetration depth with scattering in single crystals of K$_{1-x}$Na$_x$Fe$_2$As$_2$}

\author{H. Kim}
\affiliation{Ames Laboratory, Ames, IA 50010, USA}
\affiliation{Department of Physics \& Astronomy, Iowa State University, Ames, IA 50010}

\author{M. A. Tanatar}
\affiliation{Ames Laboratory, Ames, IA 50010, USA}
\affiliation{Department of Physics \& Astronomy, Iowa State University, Ames, IA 50010}

\author{Yong Liu}
\affiliation{Ames Laboratory, Ames, IA 50010, USA}

\author{Zachary Cole Sims}
\affiliation{ Department of Physics and Astronomy, The University of Tennessee, Knoxville, Tennessee 37996-1200, USA}

\author{Chenglin Zhang}
\affiliation{ Department of Physics and Astronomy, The University of Tennessee, Knoxville, Tennessee 37996-1200, USA}
\affiliation{ Department of Physics and Astronomy, Rice University, Houston, Texas 77005, USA }

\author{ Pengcheng Dai}
\affiliation{ Department of Physics and Astronomy, Rice University, Houston, Texas 77005, USA }

\author{T. A. Lograsso}
\affiliation{Ames Laboratory, Ames, IA 50010, USA}

\author{R. Prozorov}
\affiliation{Ames Laboratory, Ames, IA 50010, USA}
\affiliation{Department of Physics \& Astronomy, Iowa State University, Ames, IA 50010}

\date{16 May 2014}

\begin{abstract}

London penetration depth, $\lambda (T)$, was measured in single crystals of K$_{1-x}$Na$_x$Fe$_2$As$_2$, $x$=0 and 0.07, down to temperatures of 50~mK, $\sim T_c/50$. Isovalent substitution of Na for K significantly increases impurity scattering, with $\rho(T_c)$ rising from 0.2 to 2.2 $\mu \Omega$cm, and leads to a suppression of $T_c$ from 3.5~K to 2.8~K. At the same time, a close to $T$-linear $\Delta \lambda (T)$ in pure samples changes to almost $T^2$ in the substituted samples. The behavior never becomes exponential as expected for the accidental nodes, as opposed to $T^2$ dependence in superconductors with symmetry imposed line nodes. The superfluid density in the full temperature range follows a simple clean and dirty $d$-wave dependence, for pure and substituted samples, respectively. This result contradicts suggestions of multi-band scenarios with strongly different gap structure on four sheets of the Fermi surface.
\\This paper is published in: Phys. Rev. B \textbf{89}, 174519 (2014).
\end{abstract}


\maketitle

\section{Introduction}

The discussion of the superconducting pairing mechanism in iron-based superconductors was guided by early observations of full superconducting gap in tunneling experiments,\cite{Chen2008} which was seemingly at odds with neutron resonance peak \cite{Christianson2008} suggesting a sign change of the order parameter. Theoretically, Mazin {\it et al.} suggested pairing mechanism, in which superconducting order parameter changes sign between hole and electron bands, but each band remains fully gapped.\cite{Mazin2008,Mazin2010nature} Verification of this so-called $s_{\pm}$ pairing quickly became a focal point of studies of the superconducting gap structure.

Clear deviations from full-gap $s_{\pm}$ pairing scenario were found in nuclear magnetic resonance (NMR) and heat capacity studies of KFe$_2$As$_2$ (K122),\cite{Fukazawa2009} which represents the terminal overdoped composition of Ba$_{1-x}$K$_x$Fe$_2$As$_2$ series (BaK122).\cite{Rotter2008,Rotter2008a} Systematic doping evolution studies over the whole superconducting dome in Ba(Fe$_{1-x}$Co$_x$)$_2$As$_2$ (BaCo122),\cite{Tanatar2010kappa,Reid2010,Gordon2009prb,Gordon2010lambda0,Mu2011,JSKim2012a,JSKim2012} NaFe$_{1-x}$Co$_x$As,\cite{Cho2012Na111, Prozorov2013} and Ba$_{1-x}$K$_x$Fe$_2$As$_2$ (Ref. \onlinecite{Reid2012a}) suggest that the superconducting gap in all cases mentioned above develops evident anisotropy and even nodes at the dome edges. Thus K122 is not unique as a nodal superconductor. On the other hand, it is one of the cleanest stoichiometric materials and, therefore, understanding its superconducting gap is of great importance for the entire iron-based family.

Such diverse evolution of the superconducting gap with doping in iron-based superconductors is notably different from the cuprates, in which nodal $d$-wave pairing is observed in all doping regimes and families of materials. Several theoretical explanations of this fact were suggested.\cite{Thomale2011,Chubukov2012,Das2012} The observed doping evolution was explained in $s_{\pm}$ pairing scenario as a result of the competition between inter-band pairing and intra-band Coulomb repulsion.\cite{Chubukov2012,Glatz2010} Alternatively, it was explained to be due to a phase transition between $s_{\pm}$-wave and $d$-wave superconducting states.\cite{Thomale2011} The important difference is that the nodes in the gap structure are accidental in the former scenario but are symmetry-imposed in the latter.

The existence of line nodes in the superconducting gap of K122 is supported by a quantity of experiments. London penetration depth studies found close to $T$-linear temperature dependence.\cite{Hashimoto2010K122} The analysis of vortex lattice symmetry in small angle neutron scattering suggested horizontal line nodes in the gap.\cite{Kawano-Furukawa2011} Thermal conductivity studies revealed robust finite residual linear term in zero field, which rapidly increases with magnetic field.\cite{Dong2010,Reid2012} Moreover, residual linear term was found to be independent of the heat flow direction \cite{Reid2012} and impurity scattering\cite{Reid2012,Wang2014} suggesting presence of symmetry-imposed vertical line nodes in the superconducting gap, similar to the $d$-wave superconducting state of the cuprates.\cite{Taillefer1997} Measurements of the specific heat in Na-doped samples are consistent with a $d$-wave pairing.\cite{Abdel-Hafiez2013,Grinenko2014} Moreover, non-monotonic d
 ependence of $T_c$ on pressure was explained as an evidence of a phase transition from $d$-wave to $s$-wave symmetry  in the superconducting state of K122. \cite{Tafti2013,Fernandes2013}

However, these observations consistent with $d$-wave scenario are disputed by laser ARPES \cite{Okazaki2012} suggesting extreme multiband scenario in which the all line nodes are observed only on one hole band (octuplet  node scenario), with three other sheets being fully gapped. Two recent heat capacity studies \cite{Hardy2013,Kittaka2014}  observed a clear feature at around 0.7~K, with the general view of the curves very similar to the multi-band MgB$_2$.\cite{Bouquet} Hardy {\it et al.} \cite{Hardy2013} pointed out the importance of measurements with temperatures below 100~mK and were able to fit experimental $C_e/T$ curve in the whole superconducting region, including the feature at 0.7~K, assuming four full-gap contributions, three of which have anomalously small gaps (lilliputian gap scenario). It is important to notice though that both ARPES and heat capacity measurements probe changes induced by opening of the superconducting gap in the normal state, not of the conden
 sate itself. The former is in addition probing the states at the top layer of the sample surface, prone to modification by surface reconstruction.\cite{Sr214} Heat capacity measurement is a bulk probe, but by nonselective probing the whole sample can be affected by the presence of impurity phases. The admixture of the magnetic impurity phases was invoked for the explanation of 0.7~K features in other heat capacity studies.\cite{JSKim2011impurity,Grinenko}

In this paper we report systematic studies of the London penetration depth in pure KFe$_2$As$_2$ and isovalently substituted K$_{1-x}$Na$_x$Fe$_2$As$_2$ (KNa122). We show that the temperature-dependent superfluid density calculated with experimental London penetration depth and its response to the pair-breaking due to non-magnetic scattering are consistent with the symmetry-imposed line nodes in the superconducting gap in contrast to the extreme multi-band scenario.

\section{Experimental}

Single crystals of KFe$_2$As$_2$ were grown using the KAs flux method as explained in Ref. \onlinecite{Liu2013}. Single crystals of K$_{1-x}$Na$_x$Fe$_2$As$_2$ were grown by mixing (NaK)As/FeAs in sealed Tantalum tubes, followed by cooking at $1150 ~^\circ \textmd{C}$ for 3 hours and  $5 ~^\circ \textmd{C/hr}$ cooling down to room temperature.\cite{Kihou2010} The wavelength dispersive x-ray spectroscopy (WDS) in {\it JEOL JXA-8200} electron microprobe was utilized to determine the chemical compositions. The actual concentration $x$ was determined by averaging results of 12 measurements on different locations per single crystal, statistical error of the composition is $\pm$0.005. Small resistance contacts ($\sim$ 10 $\mu \Omega$) were tin-soldered \cite{SUST,patent}, and in-plane resistivity was measured using a four probe technique in {\it Quantum Design PPMS}. The London penetration depth was measured in samples with typical dimensions of 0.8$\times$0.8$\times$0.1 mm$^3$ by
 using a tunnel diode resonator (TDR) technique \cite{Degrift1975} in a $^3$He cryostat and a dilution refrigerator with operation frequencies of $f_0=14$ MHz and 17 MHz, respectively. The $^3$He-TDR was used for measurements down to $T=0.5$ K, and the measurements were extended to lower temperatures down to $T=0.05$ K by using the dilution refridgerator-TDR. The samples were inserted into a 2 mm inner diameter copper coil that produces an rf excitation field with amplitude $H_{ac} \sim 20$ mOe which is much smaller than typical first critical field. Measurements of the in-plane penetration depth, $\Delta \lambda(T)$, were done with $H_{ac} \parallel c$-axis. The shift of the resonant frequency is related to magnetic susceptibility of the specimen via $\Delta f(T)=-G4\pi\chi(T)$  (in cgs units) where $\chi(T)$ is the differential magnetic susceptibility, $G=f_0V_s/2V_c(1-N)$ is a constant, $N$ is the demagnetization factor, $V_s$ is the sample volume, and $V_c$ is the coil vo
 lume. The constant $G$ was determined from the full frequency change by physically removing the sample out of the coil. With the characteristic sample size, $R$, which can be calculated by the procedure explained in Ref.~\onlinecite{Prozorov2011}, $4\pi\chi=(\lambda/R)\tanh (R/\lambda)-1$, from which $\Delta \lambda$ can be obtained.\cite{Prozorov2000,Prozorov2011}
The frequency shift measured with TDR technique in the normal state represents skin depth, $\delta$, provided that dimensions of the sample are much greater than $\delta$ and are due to normal skin effect.\cite{Hardy1993,Kim2011LiFeAs} This measured skin depth can be converted into the resistivity, $\rho$, by using the relation $\delta=c/(2\pi\sqrt{\rho/f_0})$ (cgs units), where $c$ is the speed of light.

\section{Results and Discussion}

\begin{figure}
\includegraphics[width=0.95\linewidth]{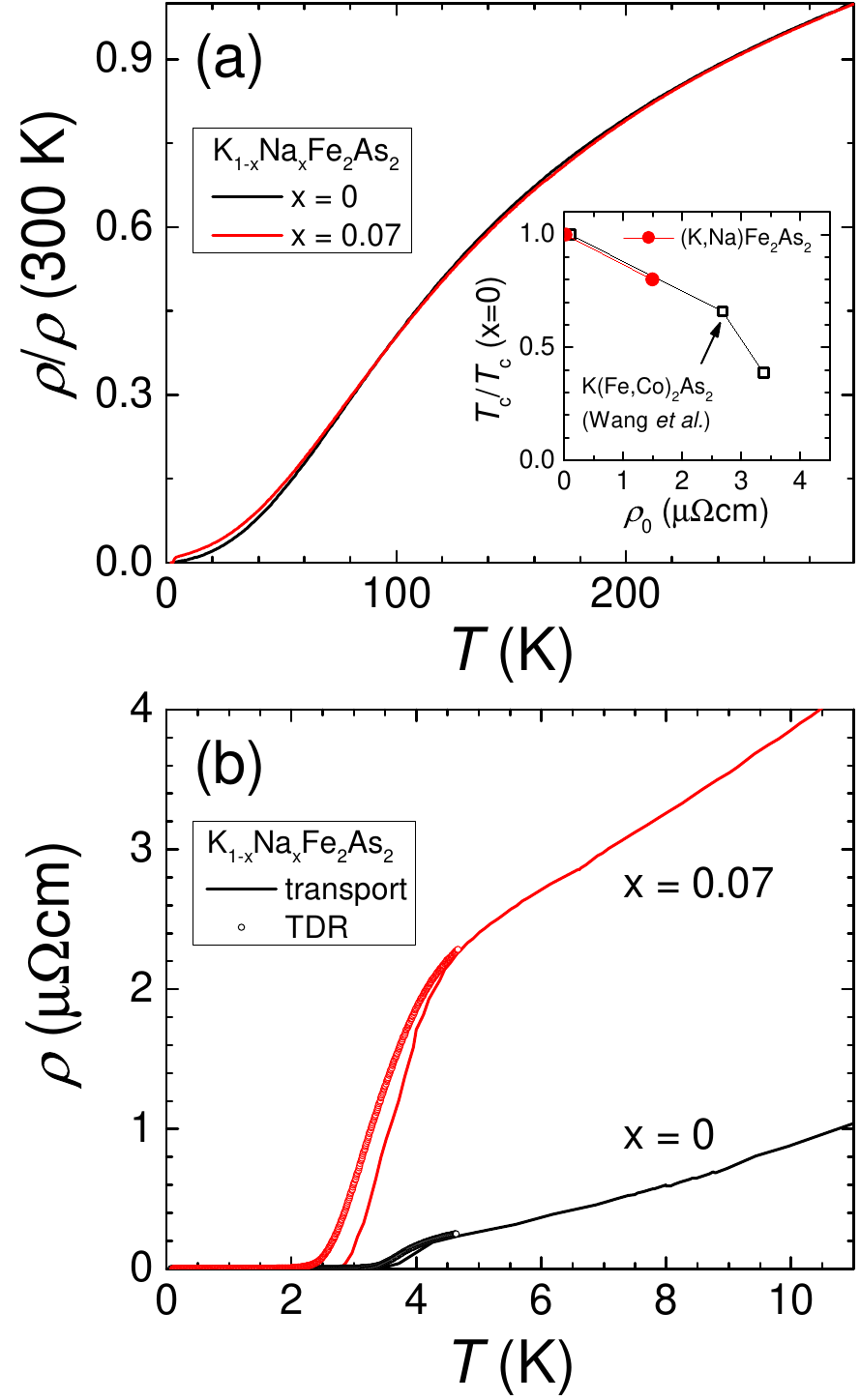}%
\caption{ (a) In - plane electrical resistivity of K$_{1-x}$Na$_x$Fe$_2$As$_2$ with $x$=0 and $x$=0.07 shown over the full temperature range using normalized resistivity scale, $\rho(T)/\rho(300K)$ and (b) zoomed on the superconducting transition region (lines) using actual $\rho$ scale. Data points in panel (b) show temperature-dependent resistivity as determined from radio frequency skin-depth measurements in our TDR set-up. Inset in panel (a) shows superconducting $T_c$, determined using zero-resistance criterion, as a function of the residual resistivity estimated from the Fermi - liquid fit, $\rho = \rho_0 + A T^2$ in isovalent-substituted K$_{1-x}$Na$_x$Fe$_2$As$_2$ (circles) in comparison with the electron-doped K(Fe$_{1-x}$Co$_x$)$_2$As$_2$ (squares) obtained by Wang \textit{et al.}.\cite{Wang2014}}
\label{fig1}
\end{figure}

Panel (a) of Fig.~\ref{fig1} shows temperature-dependent resistivity of K$_{1-x}$Na$_x$Fe$_2$As$_2$  with $x$=0 and 0.07 using the normalized resistivity scale $\rho/\rho(300K)$. Zoom on the superconducting transition in panel (b) shows the same data in actual resistivity values. Because of the significant scatter in the resistivity values in iron based superconductors due to the presence of hidden cracks,\cite{anisotropy,pseudogap} we used statistically significant $\rho(300K)=285\pm50$ $\mu\Omega$cm (as determined in Ref.~\onlinecite{Liu2014} by the average and standard deviation of measurements on 12 crystals) for pure samples. The values at the lower boundary of error bars were providing the best agreement with TDR skin depth measurements, less prone to cracks.\cite{Kyuil} Within rather big error bars of the resistivity measurements, the resistivity value for Na-doped samples is indistinguishable from that of the pure material at high temperatures, so we adopted the same
 $\rho(300K)$. The $\rho(T)$ of two sets of samples are identical a
 s well, except for the increased residual resistivity and suppression of the superconducting transition temperature in $x$=0.07 samples. The actually measured values of resistivity before the first signatures of superconductivity are 0.2 and 2.2 $\mu \Omega$cm. Because of the strong temperature dependence of resistivity before onset of the superconducting transition, these values are significantly larger that the extrapolated to $T$=0 residual resistivities of 0.100$\pm$0.050 ($x$=0) and 1.7 $\mu \Omega$cm ($x$=0.07). The skin depth measured by TDR technique show good agreement with the direct transport measurements as shown in Fig.~\ref{fig1}(b).

Scattering introduced by Na-substitution in K$_{1-x}$Na$_x$Fe$_2$As$_2$ is clearly non-magnetic, however it provides strong pair breaking, as expected in unconventional superconductors, and substantially suppresses $T_c$. The inset in Fig.~\ref{fig1}(a) shows $T_c$, as determined using $\rho$=0 criterion, as a function of $\rho_0$. For the reference we show similar data obtained in samples with aliovalent Co substitution in K(Fe$_{1-x}$Co$_x$)$_2$As$_2$.\cite{Wang2014} Despite the fact that Co-substitution provides electron-doping, while Na-substitution is isoelectronic, both types of substitution introduce similar pair-breaking, suggestive that scattering, rather than electron count, plays primary role in $T_c$ suppression.

\begin{figure}
\includegraphics[width=0.95\linewidth]{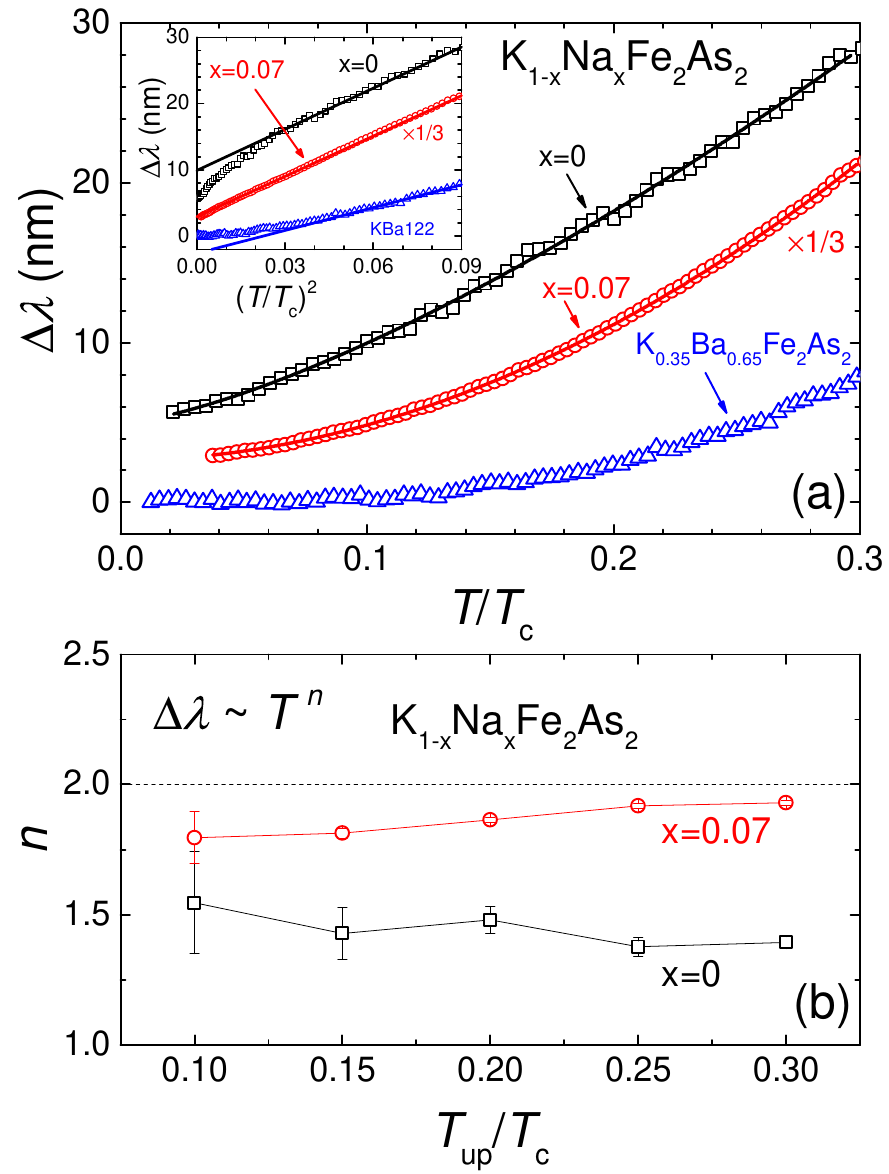}%
\caption{ (a) Low temperature ($T \leq 0.3T_c$) variation of the London penetration depth in K$_{1-x}$Na$_x$Fe$_2$As$_2$, $x$=0 and $x$=0.07, compared with a full-gap superconductor K$_{0.35}$Ba$_{0.65}$Fe$_2$As$_2$.\cite{Salovich2013} The inset shows the same data plotted vs. $(T/T_c)^2$ to highlight a sub-quadratic behavior of $\lambda(T)$ in clean samples and almost perfect $T^2$ dependence in Na-substituted samples. Data for K$_{1-x}$Na$_x$Fe$_2$As$_2$, $x=$0 and $x=$0.07 are vertically shifted for clarity by 6 and 3 nm, respectively. The bottom panel (b) shows the exponent $n$ in the power-law fits of the data, $\Delta \lambda (T) =AT^n$,  plotted versus the upper limit of the fitting range from the base temperature, $T_{min} =$ 50 mK to $T_{up}$. Note that neither $\Delta \lambda (T)$ raw data of panel (a), nor $n(T_{up})$ show any irregularities over the whole range (up to 1~K), the range where significant anomalies are observed in the specific heat data at and below 0
 .5-0.8 K.\cite{Hardy2013,Kittaka2014}}
\label{fig2}
\end{figure}


The temperature-variation of the London penetration depth provides information about the structure of the superconducting gap. This statement is valid in a characteristic temperature range for which the superconducting gap $\Delta(T)$ can be considered constant. For single gap superconductors the upper limit is approximately $T_c/3$.
Below this temperature, $\Delta \lambda (T)$ shows exponential saturation in single-gap $s$-wave superconductors. In multiband situation, however, the smallest gap determines the region of exponential behavior, and it can be much smaller than $T_c/3$. This is demonstrated in Fig.~\ref{fig2}(a) for K$_{0.35}$Ba$_{0.65}$Fe$_2$As$_2$ (Ref. \onlinecite{Salovich2013}) where the ``upper low-temperature range'' extends only up to $T_c/6$. Therefore, lowest temperature experiments are needed to probe multiband superconductivity.

For the gap with symmetry-imposed line nodes, $\Delta\lambda$ can be described by the power-law, $\Delta\lambda(T)=AT^n$, where the exponent $n$ depends on impurity scattering and symmetry of the order parameter. For line nodes, the exponent $n$ \textit{increases} from the value of $n=1$ in the clean limit to $n=2$ in the dirty limit.\cite{Hirschfeld1993} In $s_{\pm}$ pairing, however, the exponential dependence is expected for a clean case. Experimentally the data are still analyzed in terms of the power-law and for $n$ greater than 3 it is difficult to distinguish from the exponential with a realistic noise of the data points.
Adding scattering in the $s_\pm$ pairing case, the exponent $n$ {\it decreases} from a larger value to the same dirty limit value of $n=$2.\cite{Vorontsov2009} However, in the case of $s_{\pm}$ pairing with accidental nodes, disorder will lift the nodes resulting in a change of the exponent $n$ from 1 to exponential.\cite{Mishra2009}

In Fig.~\ref{fig2}(a) we show $\Delta \lambda$ vs. $t=T/T_c$ of KNa122 crystals with $x$=0 and $x$=0.07. Inset shows the same data plotted as $\Delta \lambda$ vs. $t^2$. It is clear that the exponent $n$ \emph{increases} with doping and, in the $x=$0.07 sample, approaches the dirty-limit value $n$=2. Both features are consistent with the superconducting gap with the symmetry-imposed line nodes. Specifically, the data between the base temperature of 50 mK and $T=T_c/3$ can be best fitted to the power-law function with $n=1.39$, $A=200$ nm and $n=1.93$, $A=911$ nm for $x=$0 and 0.07, respectively. Both are in the range expected for symmetry imposed line nodes. According to Hirschfeld-Goldenfeld's theory,\cite{Hirschfeld1993} the penetration depth can be interpolated as $\Delta \lambda(T)=\lambda(0)T^2/(T+T^*)$ where $T^*$ is a crossover temperature from $T$ to $T^2$ behavior at the low temperatures. Our fit in pure K122 using this formula gives $T^*=0.5T_c$. On the other hand,
 this expression is inapplicable for Na-substituted sample that shows quadratic behavior at all temperatures indicating that these samples are in the dirty limit.

Alternatively, as discussed above, this crossover behavior can be due to multi-band effects in superconductivity. For multi-band superconductors the upper end of the characteristic $\Delta \lambda (T)$ dependence is determined by the smaller gap $\Delta _{min}$, and shrinks proportional to $\Delta_ {min}/\Delta$. Since gap structure is not known, the upper limit of the characteristic behavior cannot be assumed {\it a priori}. Therefore, for quantitative analysis of the data we performed power-law fitting over the variable temperature interval. The low-temperature end of this interval was always kept fixed at the base temperature of about 50 mK, and the exponent of the power-law fit $n$ was determined as a function of the high-temperature end, $T_{up}$. This dependence of $n(T_{up})$ for KNa122 samples with $x$=0 and $x$=0.07 is shown in Fig.~\ref{fig2}(b). Note that neither raw $\Delta \lambda (T)$ data of Fig.~\ref{fig2}(a), nor $n(T_{up})$ of Fig.~\ref{fig2}(b), shows any i
 rregularities over the range $t \leq 0.3$, which is consistent with other penetration depth measurements. \cite{Hashimoto2010K122,Ohishi2012} But this is in stark contrast with two reports of significant anomalies at about 0.7~K and below in the temperature-dependent specific heat.\cite{Hardy2013,Kittaka2014}


\begin{figure}
\includegraphics[width=0.95\linewidth]{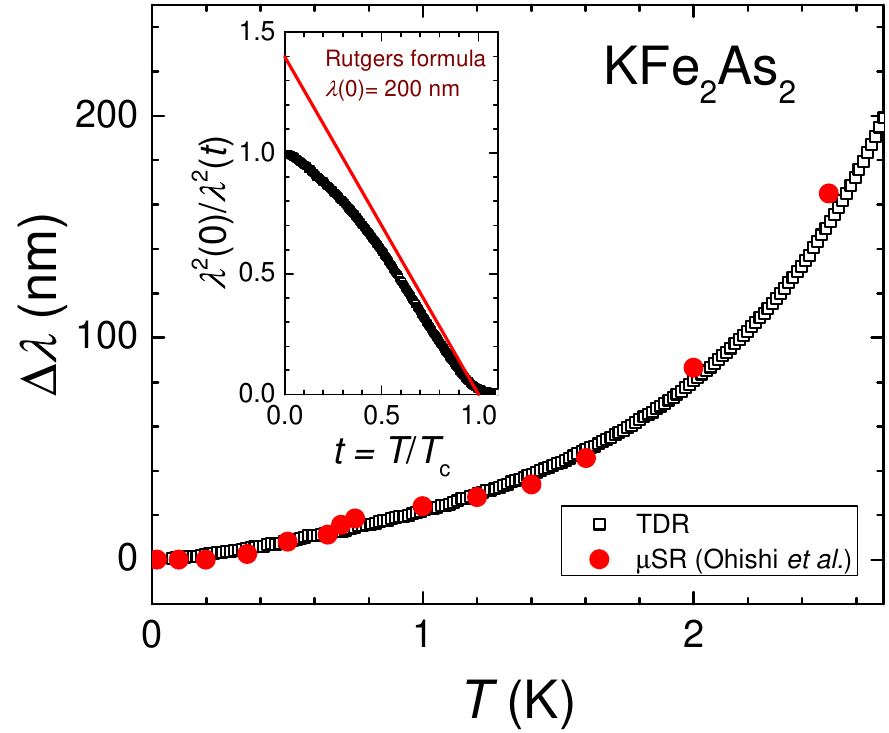}%
\caption{ (Color online) Variation of the London penetration depth in pure KFe$_2$As$_2$ (black squares) compared to the results from muon spin rotation experiments (red circles) by Ohishi \textit{et al.}.\cite{Ohishi2012} The ($\mu$SR) data provide zero-temperature value of $\lambda(0)$=200~nm, which we use to calculate the superfluid density $\rho_s(T)$, as shown in the inset. The line in the inset shows a slope of the $\rho_s(T)$ curve at $T_c$ calculated using thermodynamic Rutgers relation\cite{Kim2013a} from the specific heat jump and the slope of the $H_{c2}(T)$ at $T_c$.}
\label{fig3}
\end{figure}

A further insight into the structure of the superconducting gap in KFe$_2$As$_2$ can be obtained through the analysis of the temperature-dependent superfluid density, $\rho_s(T)=\lambda^2(0)/\lambda^2(T)$, with $\lambda (T)=\lambda(0)+\Delta \lambda (T)$.
This quantity can be calculated for known superconducting gap structure and compared with the experiment over the full superconducting temperature range. To perform this analysis the knowledge of $\lambda(0)$ is required, which is not readily available from our experiments. We used the value of $\lambda(0)$=200 nm based on a recent muon spin relaxation $\mu$SR experimental results\cite{Ohishi2012} as well as experimental plasma frequencies\cite{Hashimoto2010K122} and, we verify its consistency by using thermodynamic Rutgers formula.\cite{Kim2013a} We note that 200 nm is the typical value for iron based superconductors away from the coexistence with magnetism regime.\cite{Gordon2010lambda0,Li2008,Hashimoto2012QCP}

\begin{figure}
\includegraphics[width=1\linewidth]{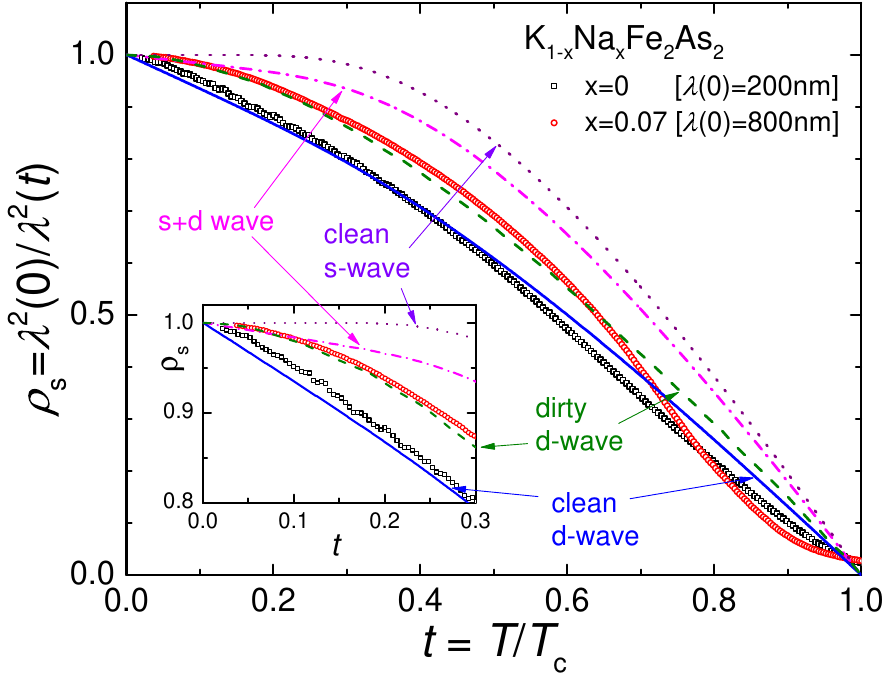}%
\caption{ (Color online) Superfluid density for K$_{1-x}$Na$_x$Fe$_2$As$_2$ $x=$0 and $x=$0.07 compared with the theoretical predictions for different models of the superconducting gap structure. We include the single band clean limit $s$-wave (dotted line) and $d$-wave (full line) cases, as well as dirty limit $d$-wave (dashed line). We also show superfluid density for multiband scenario with nodes only on some of the bands (shown as ``$s+d$'' by dot-dashed line), calculated using parameters of the Fermi surface and a combination of nodal and nodeless gaps as suggested by Okazaki {\it et al.} (Ref.~\onlinecite{Okazaki2012}). The data for pure compound follow closely a clean limit $d$-wave curve, whereas the $\rho_s(T)$ for sample $x=0.07$ follow the dirty $d$-wave dependence in substantial portion of the full temperature range. The inset zooms at the low-temperature part highlighting the differences in behavior.}
\label{fig4}
\end{figure}

The main panel of Fig.~\ref{fig3} compares our TDR data (black squares) with $\mu$SR data (red dots) showing an excellent agreement. The inset shows superfluid density calculated using $\lambda(0)$=200 nm and comparison with the expected slope, calculated using the thermodynamic Rutgers formula \cite{Kim2013a} that connects the slope of $\rho_s(T)$ at $T_c$, $\rho_s^{\prime} \equiv \frac{d \rho_s}{dT}$, with the slope of the $H_{c2}(T)$ at $T_c$, $H_{c2}^{\prime} \equiv \frac{d H_{c2}}{dT}$, and the magnitude of the specific heat jump $\Delta C$ at $T_c$ via:

\begin{equation}
\frac{\rho_s^{\prime}(T_c)}{\lambda^2(0)} = \frac{16 \pi^2 T_c \Delta C}  {\phi_0 H_{c2}^{\prime}(T_c)}  \,,
\label{Rutgers}
\end{equation}

\noindent where $\phi_0$ is magnetic flux quantum. Taking the slope $H_{c2}^{\prime}$=-0.55 T/K (Ref. \onlinecite{Abdel-Hafiez2012}) and specific heat jump as $\Delta C$= 159.6 mJ/mol K (Ref. \onlinecite{Abdel-Hafiez2012}) we obtain the slope $\rho_s^{\prime}$=-1.4 as shown in the inset in Fig.~\ref{fig3}. It can be seen that the value of $\lambda(0)$=200 nm is quite reasonable. For the sample with $x$=0.07 we estimated $\lambda(0)=0.8$ $\mu$m  using Homes scaling based on the resistivity and $T_c$ data.\cite{Homes2004,Kogan2013} The resulting $\rho_s(T)$ for both pure and $x$=0.07 samples are shown in Fig.~\ref{fig4}. The low-temperature parts of $\rho_s(T)$ are zoomed in the inset of Fig.~\ref{fig4}. For comparison, Fig.~\ref{fig4} shows expected temperature dependent superfluid density in clean (full line) and dirty (dashed line) $d$-wave cases, which are representative of all superconductors with symmetry-imposed line nodes. Clearly, in the large portion of the full tempe
 rature range, the data for pure K122 follow a simple $d$-wave behavior, whereas the Na-substituted sample follows expectations for a dirty $d$-wave superconductor.

Since multi-band scenario is clearly suggested by both ARPES \cite{Okazaki2012} and heat capacity studies \cite{Hardy2013,Kittaka2014} we now turn to a more realistic band/superconducting gap structure. Here we try to estimate the temperature-dependent superfluid density from the published band-structure, ARPES, quantum oscillations, and specific heat data, and we compare it with the experimental results. Three hole-like sheets of the Fermi surface centered around the $\Gamma$-point will be considered. ARPES measurements concluded that there are two full, somewhat anisotropic, gaps on the inner, $\alpha$, ($\Delta/k_{B}T_{c}$ varies between 2.7 and 5) and the outer, $\beta$, ($\Delta/k_{B}T_{c}$ varies between 0.4 and 0.6) sheets and a nodal gap on the middle ($\zeta$) sheet ($\Delta/k_{B}T_{c}$ varies between 0 and 2.6).\cite{Okazaki2012} In sharp contrast, the gap amplitudes from the fit of the specific heat are $0.57,$ $0.22,$ $0.35$ and the surprisingly large, $1.90$, gap
  on the electron-like $\epsilon$ Fermi sheet near $X$ points.\cite{Hardy2013}

In a multiband system different sheets of the Fermi surface contribute partial superfluid densities as:%
\[
\rho\left(  t\right)  =\sum\gamma_{i}\rho_{i}\left(  t\right)
\]
where the sum is ran over all contributing sheets and
\[
\gamma_{i}=\frac{n_{i}v_{i}^{2}}{\sum n_{i}v_{i}^{2}}%
\]
where%
\[
n_{i}=\frac{N_{i}\left(  0\right)  }{N\left(  0\right)  }=\frac{N_{i}\left(
0\right)  }{\sum N_{i}\left(  0\right)  }%
\]
is the normalized total density of states at each band for both spin
directions and $v_{i}$ is the Fermi velocity. To evaluate superfluid density and to estimate the $\gamma_{i}$ factors it is convenient to use plasma frequency via:
\[
\frac{1}{\lambda^{2}\left(  0\right)  }=\frac{8\pi e^{2}N\left(  0\right)
\sum n_{i}v_{i}^{2}}{3c^{2}}=\frac{\left\langle \omega_{p}^{2}\right\rangle
}{c^{2}}%
\]
so knowing partial $\omega_{pi}^{2}$ we can express $\gamma_{i}$ as:%
\[
\gamma_{i}=\frac{\omega_{pi}^{2}}{\sum\omega_{pi}^{2}}%
\]
The partial and total plasma frequencies were reported from DFT calculations,\cite{Hashimoto2010K122} giving $\gamma_{1}=0.71$ and $\gamma_{1}=0.29$ for a full and for a nodal gap, respectively, and $\gamma _{1}=0.77$ and $\gamma_{1}=0.23$ from de Haas-van Alphen measurements.\cite{Terashima2010} In case of specific heat analysis, no Fermi velocities or plasma frequencies are reported and we can only use pure 2D approximation where $nv^{2}=k_{F}^{2}/\pi m=k_{F}^{2}/\hbar^{2}\pi^{2}N(0)$. Reported densities of states are roughly the same, so the contribution to the superfluid density depends on the Fermi wave vector and is roughly a quarter for the middle band, consistent with the above numbers.

Although the modulation of the gaps might play some role in determining the temperature dependence of the superfluid density, and the gaps must satisfy a self-consistency relation in a multiband system, the largest gap will always have the BCS-like temperature variation. At least in the case of specific heat data analysis, the gaps were calculated self-consistently and they, indeed, confirm the above statement. We therefore can compare two scenarios: one which mimics ARPES and specific heat findings where there are two effective gaps: nodeless and nodal (with different partial densities of states for two different experiments) and the alternative when the gaps possess $d$-wave everywhere. The latter must be true for all hole-like sheets of the Fermi surface if the pairing potential changes sign along the diagonals of the Brillouin zone. In that case the normalized superfluid density will be just a simple single-gap $d$-wave.

When analyzing $\rho_s (T)$, we should notice that deviation from $d$-wave calculations do not leave much room for any full gap contribution to the superfluid density. If it was present, at a level more that 0.1 of the total $\rho_s$, it would result in significant exceeding of $\rho_s(T)$ over the curve for a $d$-wave case. Based on this comparison we can disregard any contributions from full gap-superfluid in both clean $x$=0 and $x$=0.07 samples with the accuracy of less than 0.1 of the total superfluid density.

\section{Conclusion}

Along with the power-law behavior of $\Delta\lambda(T)\sim T^{1.4}$ at low temperatures, the temperature response of the superfluid in both clean and dirty samples is not only consistent with the existence of line nodes in the superconducting gap, but does not leave much room for any contribution from Fermi surfaces with a large and dominant full gap, as suggested by ARPES\cite{Okazaki2012} and heat capacity.\cite{Hardy2013}

In conclusion, our resistivity and TDR London penetration depth studies on high quality pure and isoelectron Na-substituted KFe$_2$As$_2$ find the behavior which is consistent with the  expectations for a superconductor with symmetry-imposed line nodes.

\section{Acknowledgements}

We thank W.~E.~Straszheim for performing WDS analysis and P. Hirschfeld, A. Chubukov and R.~M.~Fernandes for useful discussions. This work was supported by the U.S. Department of Energy (DOE), Office of Science, Basic Energy Sciences, Materials Science and Engineering Division. The research was performed at the Ames Laboratory, which is operated for the U.S. DOE by Iowa State University under contract DE-AC02-07CH11358. The single crystal growth at the University of Tennessee and Rice University was supported by U.S. DOE, BES under Grant No. DE-FG02-05ER4620 (P.D.).


\end{document}